# Pairing Interaction and Self-Consistent Densities in Neutron-Rich Nuclei

J. Dobaczewski,[1,2,3] W. Nazarewicz,[1,4,5] and P.-G. Reinhard[3,6]

[1] *Institute of Theoretical Physics, Warsaw University, ul. Hoża 69, PL-00681, Warsaw, Poland*
[2] *Joint Institute for Heavy Ion Research, Oak Ridge National Laboratory, P.O. Box 2008, Oak Ridge, Tennessee 37831*
[3] *Institute for Nuclear Theory, University of Washington, Seattle, Washington 98195*
[4] *Department of Physics and Astronomy, University of Tennessee, Knoxville, Tennessee 37996*
[5] *Physics Division, Oak Ridge National Laboratory, P.O. Box 2008, Oak Ridge, Tennessee 37831*
[6] *Institut für Theoretische Physik, Universität Erlangen, Staudtstr. 7, D-91058 Erlangen, Germany*

Particle and pairing densities in spherical even-even neutron-rich nuclei are studied within the Skyrme-Hartree-Fock-Bogoliubov approach with the density-dependent pairing interaction. The influence of the density dependence of the pairing interaction on asymptotic properties of nucleonic distributions are analyzed. It is demonstrated that the size of the neutron halo dramatically depends on the behavior of the pairing interaction at low density.

## I. INTRODUCTION

One of the main avenues addressed by radioactive ion beams is the evolution of nuclear structure as a function of neutron-to-proton asymmetry. From a theoretical point of view, exotic nuclei far from stability offer a unique test of those components of effective interactions that depend on the isospin degrees of freedom. Since effective interactions in heavy nuclei have been adjusted to stable nuclei and to selected properties of infinite nuclear matter, it is by no means obvious that the isotopic trends far from stability, predicted by commonly used effective interactions, are correct. Such investigations reach inevitably into the regime of weakly bound nuclei, and these are much more difficult to treat theoretically than well-bound systems [1]. Their Fermi energy lies very close to zero, and the decay channels must be taken into account explicitly. Correlations due to pairing, core polarization, and clustering become crucial. In a dripline system, the pairing interaction and the presence of skin excitations (soft modes) could require going beyond the picture of a nucleon moving in a single-particle orbit [2–6]. The positive aspect of this situation is that dripline nuclei are also critical probes to understand and to further develop the nuclear pairing force.

Surprisingly, rather little is known about the basic properties of the pairing force. Up to now, the microscopic theory of the pairing interaction has only seldom been applied in realistic calculations for finite nuclei. A "first-principle" derivation of the pairing interaction from the bare $NN$ force still encounters many problems such as, e.g., treatment of core polarization [7,8]. Hence, phenomenological pairing interactions are usually introduced. In most older nuclear structure calculations, the pairing Hamiltonian has been approximated by the state-independent seniority pairing force, or schematic multipole pairing interaction [9]. Such oversimplified forces, usually treated by means of the BCS approximation, perform remarkably well when applied to nuclei in the neighborhood of the stability valley, but they are inappropriate (and formally wrong) when extrapolating far from stability. The self-consistent mean-field models have meanwhile reached such a high level of precision that one needed to improve on the pairing part of the model. The presently most up-to-date models employ local paring forces parametrized as contact interactions [10,11,2]. More flexible forms attach a density dependence to the pairing strength [12–14]. There exist even more elaborate forms which include also gradient terms [15,16]. It is clear that some density dependence is needed. Nuclear matter calculations and experimental data on isotope shifts strongly suggest that pairing is a surface phenomenon and that the pairing interaction should be maximal in the surface region. It is not so obvious, however, how the actual density dependence should be parametrized in detail [2]. This is why neutron-rich nuclei play such an important role in this discussion. Indeed, because of strong surface effects, the properties of these nuclei are sensitive to the density dependence of pairing. Last, but not least, a good understanding of pairing is important for astrophysical applications where exotic nuclei play a crucial role and where the nuclear pairing interaction is also important for theories of superfluidity in neutron stars [17].

The main objective of this work is to investigate the role of density dependence of pairing interaction on properties of neutron-rich nuclei. The material contained in this study is organized as follows. The definitions pertaining to self-consistent densities and mean fields and a specification of interactions are given in Sec. II. The analysis of HFB densities and mean-field potentials obtained in different pairing models is given in Sec. III. The results of self-consistent calculations for pairing gaps, separation energies, and halos are discussed in Sec. IV. Finally, Sec. V contains the main conclusions of this work.

## II. HARTREE-FOCK-BOGOLIUBOV DENSITIES AND MEAN FIELDS

This section contains a very brief description of the HFB formalism in the coordinate representation. Since the method is standard, our discussion is limited to the most essential definitions and references. For more details, we would like to refer the reader to the previous work [18,19,2].

The HFB approach is a variational method which uses nonrelativistic independent-quasiparticle states as trial wave functions [9]. The total binding energy of a nucleus is obtained self-consistently from the energy functional [20]:

$$\mathcal{E} = \mathcal{E}_{\text{kin}} + \mathcal{E}_{Sk} + \mathcal{E}_C + \mathcal{E}_{\text{pair}}, \quad (1)$$

where $\mathcal{E}_{\text{kin}}$ is the kinetic energy functional, $\mathcal{E}_{Sk}$ is the Skyrme functional, $\mathcal{E}_C$ is the Coulomb energy (including the exchange term in the Slater approximation), and $\mathcal{E}_{\text{pair}}$ is the pairing energy.

By minimizing the functional (1), one arrives at the HFB equation [18,19,2] for the two-component single-quasiparticle HFB wave function $\{\phi_1, \phi_2\}$:

$$\int d^3 r' \sum_{\sigma'} \begin{pmatrix} h(r\sigma, r'\sigma') & \tilde{h}(r\sigma, r'\sigma') \\ \tilde{h}(r\sigma, r'\sigma') & -h(r\sigma, r'\sigma') \end{pmatrix} \begin{pmatrix} \phi_1(E, r'\sigma') \\ \phi_2(E, r'\sigma') \end{pmatrix}$$
$$= \begin{pmatrix} E + \lambda & 0 \\ 0 & E - \lambda \end{pmatrix} \begin{pmatrix} \phi_1(E, r\sigma) \\ \phi_2(E, r\sigma) \end{pmatrix}, \quad (2)$$

where $\lambda$ is the Fermi energy and $h$ and $\tilde{h}$ are the particle-hole (p-h) and particle-particle (p-p) mean-field Hamiltonians.

Properties of the HFB equation in the spatial coordinates, Eq. (2), and the asymptotic properties of HFB wave functions and density distributions have been analyzed in Refs. [18,19,2]. In particular, it has been shown that the spectrum of quasi-particle energies $E$ is continuous for $|E| > -\lambda$ and discrete for $|E| < -\lambda$. Since for $\lambda < 0$ (bound system) the lower components $\phi_2(E, r\sigma)$ are localized functions of $r$, the particle and pairing density matrices,

$$\rho(r\sigma, r'\sigma') = \sum_{0 < E_n < -\lambda} \phi_2(E_n, r\sigma)\phi_2^*(E_n, r'\sigma')$$
$$+ \int_{-\lambda}^{\infty} dn(E) \phi_2(E, r\sigma)\phi_2^*(E, r'\sigma'), \quad (3a)$$

$$\tilde{\rho}(r\sigma, r'\sigma') = - \sum_{0 < E_n < -\lambda} \phi_2(E_n, r\sigma)\phi_1^*(E_n, r'\sigma')$$
$$- \int_{-\lambda}^{\infty} dn(E) \phi_2(E, r\sigma)\phi_1^*(E, r'\sigma'), \quad (3b)$$

are localized as well. For the case of a discretized continuum, the integral over the energy reduces to a discrete sum [19].

The p-h mean-field Hamiltonian can be expressed through the kinetic energy and the p-h mean-field potential:

$$h(r\sigma, r'\sigma') = T(r, r')\delta_{\sigma,\sigma'} + \Gamma(r\sigma, r'\sigma'), \quad (4)$$

where

$$\Gamma(r\sigma, r'\sigma') = \int d^3 r_2 d^3 r_2' \sum_{\sigma_2 \sigma_2'} V(r\sigma, r_2\sigma_2; r'\sigma', r_2'\sigma_2') \times$$
$$\rho(r_2'\sigma_2', r_2\sigma_2) \quad (5)$$

and $V$ is the two-body density-dependent effective p-h interaction (in our case it is a contact Skyrme interaction). The pairing mean-field potential can be expressed through the p-p density

$$\tilde{h}(r\sigma, r'\sigma') = \int d^3 r_1' d^3 r_2' \sum_{\sigma_1' \sigma_2'} 2\sigma' \sigma_2' \times$$
$$V_{\text{pair}}(r\sigma, r', -\sigma'; r_1'\sigma_1', r_2', -\sigma_2')\tilde{\rho}(r_1'\sigma_1', r_2'\sigma_2'), \quad (6)$$

where $V_{\text{pair}}$ is the two-body density-dependent effective p-p interaction. (If one adopts the philosophy of the energy density functional theory, $V$ and $V_{\text{pair}}$ can be different since they result from different variations of the energy functional.)

The local HFB densities discussed in this work,

$$\rho(r) = \sum_\sigma \rho(r\sigma, r\sigma), \quad (7a)$$

$$\tilde{\rho}(r) = \sum_\sigma \tilde{\rho}(r\sigma, r\sigma), \quad (7b)$$

have a simple physical interpretation [2]. Namely, $\rho(r)$ represents the probability density of finding a particle at a given point. On the other hand, $|\tilde{\rho}|^2(r)$ is the probability of finding a pair of nucleons in excess of the probability of finding two uncorrelated nucleons.

In this work, in the p-h channel the SLy4 Skyrme parametrization [21] has been used. This force performs well for the total energies, radii, and moments, and it is also reliable when it comes to predictions of long isotopic sequences [21].

In the particle-particle (p-p) channel, we employ the density-dependent contact interaction. As discussed in a number of papers, see e.g. Refs. [10,11,2], the presence of the density dependence in the pairing channel has consequences for the spatial properties of pairing densities and fields. The commonly used density-independent contact delta interaction, $V_{\text{pair}}^\delta(r, r') = V_0 \delta(r - r')$, leads to volume pairing. A simple modification of that force is the density-dependent delta interaction (DDDI) [12–14]

$$V_{\text{pair}}^{\delta\rho}(r, r') = f_{\text{pair}}(r)\delta(r - r'), \quad (8)$$

where the pairing-strength factor is

$$f_{\text{pair}}(r) = V_0 \left\{ 1 - [\rho_{\text{IS}}(r)/\rho_c]^\alpha \right\} \quad (9)$$

and $V_0$, $\rho_c$, and $\alpha$ are constants. In Eq. (9) $\rho_{\text{IS}}(\boldsymbol{r})$ stands for the isoscalar single-particle density $\rho_{\text{IS}}(\boldsymbol{r})=\rho_n(\boldsymbol{r})+\rho_p(\boldsymbol{r})$. If $\rho_c$ is chosen such that it is close to the saturation density, $\rho_c \approx \rho_{\text{IS}}(\boldsymbol{r}=0)$, both the resulting pair density and the pairing potential $\tilde{h}(\boldsymbol{r})$ are small in the nuclear interior, and the pairing field becomes surface-peaked. By varying the magnitude of the density-dependent term, the transition from volume pairing to surface pairing can be probed. A similar form of DDDI, also containing the density gradient term, has been used in Refs. [15,16].

Apart from rendering the pairing weak in the interior, the specific functional dependence on $\rho_{\text{IS}}$ used in Eq. (9) is not motivated by any compelling theoretical arguments or calculations. In particular, values of power $\alpha$ were chosen *ad hoc* to be either equal to 1 (based on simplicity), see e.g. Refs. [22,23], or equal to the power $\gamma$ of the Skyrme-force density dependence in the p-h channel [11,2]. The dependence of results on $\alpha$ was, in fact, never studied. In the present paper, we perform such an analysis by choosing four values of $\alpha=1$, 1/2, 1/3, and 1/6 that cover the range of values of $\gamma$ used typically for the Skyrme forces.

Calculations which are based on the contact force, such as that of Eq. (8), require a finite space of states in the p-p channel. Usually, one takes a limited configuration space determined by a cut-off in the single-particle energy or in the single-quasiparticle energy. In this work, we made a cut-off with respect to "equivalent-spectrum" single-particle energies obtained from HFB quasiparticle energies and occupation coefficients, as defined in Ref. [19]. All the quasiparticle states with $j \leq j_{\max}=21/2$ and equivalent energies up to $\bar{\epsilon}_{\max}=60$ MeV were considered, and the HFB equations were solved in the spherical box of $R_{\text{box}}=30$ fm.

This procedure differs from the prescription applied in Ref. [19] where a $(j\ell)$-dependent cut-off was used. There, the quasi-particle states with quasiparticle energies up to the depth of the effective potential $D_{j\ell}=-\min_r U_{\text{eff}}(r;j\ell)$ were considered, and in addition, at least one quasi-particle state was considered in each $(j\ell)$ block. The combination of these two rules, together with the value of $R_{\text{box}}=20$ fm used there, ensured that low-lying high-$j$ resonances were always included in the phase space. However, systematic calculations across the complete table of nuclides revealed several pathological cases where a higher-lying resonance was missing from the included phase space. This could be especially true for some near-the-barrier $j=l-1/2$ resonances, for which the effective potentials $U_{\text{eff}}$ may have small depths. Since in the present study we aim at a better description of the continuum phase space, we increased the size of the box to $R_{\text{box}}=30$ fm and changed the cut-off prescription so as to include sufficiently many states and to never miss a resonance. (For more discussion pertaining to the energy cut-off problem, see Appendix B of Ref. [2].)

For $\rho_c$ we took the standard value of 0.16 fm$^{-3}$, and the strength $V_0$ of DDDI was adjusted according to the prescription given in Ref. [11], i.e., so as to obtain in each case the value of 1.256 MeV for the average neutron gap in $^{120}$Sn. For $\alpha=1$, 1/2, 1/3, and 1/6 the adjusted values are $V_0=-520.5$, $-787.7$, $-1041.3$, and $-1772.5$ Mev fm$^{-6}$, respectively. The resulting pairing-strength factors (9) are shown in Fig. 1 as functions of density $\rho$ for the four values of the exponent $\alpha$. It is seen that for $\rho \gtrsim 0.04$ fm$^{-3}$ the pairing-strength factor $f_{\text{pair}}$ is almost independent of the power $\alpha$. At low densities, however, the pairing interaction becomes strongly dependent on $\alpha$ and very attractive at $\rho \to 0$. The pattern shown in Fig. 1 indicates that pairing forces characterized by small values of $\alpha$ should give rise to pair fields peaked at, or even beyond, the nuclear surface (halo region) where the nucleonic density is low.

### III. LOCAL HFB DENSITIES AND MEAN-FIELD POTENTIALS IN NEUTRON-RICH NUCLEI

Local p-h and p-p densities are basic elements of the HFB-Skyrme theory: they determine self-consistent fields; hence the static properties of the nucleus such as the binding energy, radius, and shape. The particle and pairing local HFB+SLy4 neutron densities $\rho_n(r)$ and $\tilde{\rho}_n(r)$ calculated for several values of $\alpha$ are displayed in Fig. 2 for $^{150}$Sn.

With decreasing $\alpha$, the p-p density $\tilde{\rho}_n(r)$ develops a long tail extending towards large distances. This is a direct consequence of the attractiveness of DDDI at low densities when $\alpha$ is small. While in the nuclear interior, the p-h density $\rho_n(r)$ depends extremely weakly on the actual form of the pairing interaction; its asymptotic values are significantly increased when $\alpha$ gets small (see inset). Moreover, one observes a clear development of a halo structure, i.e., a smooth exponential decrease, that for $\alpha=1$ starts at $r \simeq 6$ fm, for small $\alpha$ is interrupted at $r \simeq 9$ fm, and replaced by a significantly slower decrease of the density.

This is a direct consequence of the self-consistent coupling between p-h and p-p parts of the HFB Hamiltonian. That is, the increased probability of finding a correlated pair of neutrons at large distances impacts the probability of finding a single neutron in the halo region. It is to be noted that the 'halo' effect seen for $\alpha=1/6$ solely results from pairing, and it is not related to reduced binding energy. In contrast, as it is shown below, increased pairing correlations lead to greater separation energies and lower chemical potentials, i.e., to increased particle stability.

As emphasized in Ref. [2], pairing correlations can be enhanced in weakly bound nuclei due to increased surface effects and the closeness of the particle continuum. In turn, pairing can influence quite dramatically the asymptotic properties of density distributions in drip-line systems. This is nicely illustrated in Fig. 3 which compares the HFB+SLy4 densities calculated with $\alpha=1/2$ for $^{120}$Sn (well bound), $^{150}$Sn (weakly bound), and $^{170}$Sn (very weakly bound) drip-line nuclei. One can see that adding neutrons results in a simultaneous increase of density both in the nuclear interior and in the surface region. At very large distances, the asymptotic behavior of $\rho_n(r)$ reflects the gradual rise of the Fermi level $\lambda$ with a neutron number. However, this effect is much stronger for the pairing density [2]. Indeed, as seen in Fig. 3, as compared to $^{120}$Sn, there is a dramatic increase in $\tilde{\rho}_n(r)$ in the outer regions of weakly bound nuclei $^{150}$Sn and, in particular, $^{170}$Sn. These calculations indicate that for small values of $\alpha$ the box size should be chosen as very large if one aims for a very accurate description of HFB densities at large distances.

The structure of HFB densities determines the behavior of the self-consistent p-h and p-p potentials. Figure 4 shows the behavior of $U(r) \equiv \Gamma(r,r)$ (5) (local part only - see discussion in Ref. [2]) and $\tilde{U}(r) \equiv \tilde{h}(r,r)$ (6) obtained for neutrons in the weakly bound nucleus $^{150}$Sn in the HFB+SLy4 model for several values of $\alpha$. The behavior of the pairing potential $\tilde{U}(r)$ is consistent with the pattern shown in Fig. 1. Indeed, for DDDI, the pairing potential is proportional to the product of the pairing density $\tilde{\rho}$ and the pairing strength factor $f_{\text{pair}}$ (9). Consequently, $\tilde{U}(r)$ is essentially peaked around the nuclear surface, and both its minimum and range shift towards larger values of $r$ with decreasing $\alpha$. For $\alpha=1/6$, the pairing potential is still sizeable at large distances reaching 14 fm (i.e., twice the nuclear radius). The central p-h potential $U(r)$ very weakly depends on the form of the pairing interaction. It is only for small values of $\alpha$ that, due to a direct contribution from the pairing density to the p-h potential [see Eq. (A.5a) in Ref. [19]], a small barrier develops just beyond the nuclear surface. That is, the central neutron potential becomes slightly *repulsive* at $r \simeq 9$ fm. However, this effect is more than compensated by the increased pairing potential and the total binding energy decreases.

Figure 5 compares the HFB+SLy4 HFB potentials calculated with $\alpha=1/2$ for $^{120,150,170}$Sn. With increasing neutron number, the radius of the p-h potential increases, and the potential becomes more wide in the outer region (i.e., it becomes more diffused). The p-p potential becomes more surface-peaked and its range increases. By analyzing Figs. 4 and 5, one can conclude that it is in $N$-rich weakly bound nuclei that the density dependence of the pairing interaction is most important.

## IV. PAIRING GAPS, SEPARATION ENERGIES, AND HALOS

A key feature of the nucleonic density $\rho(\boldsymbol{r})$ is its second radial moment, i.e., the rms radius $R_{\text{rms}} = \langle r^2 \rangle^{1/2}$. We consider here and in the following the scaled rms radius,

$$R_{\text{geom}} = \sqrt{\frac{5}{3}} R_{\text{rms}} = \sqrt{\frac{5}{3}} \sqrt{\frac{\int d^3\boldsymbol{r}\, r^2 \rho(\boldsymbol{r})}{\int d^3\boldsymbol{r}\, \rho(\boldsymbol{r})}}, \qquad (10)$$

which we call the geometric radius [24]. The factor $(5/3)^{1/2}$ serves to bring that value closer to the box radius. Further density characteristics are best deduced from the corresponding form factor

$$F(\boldsymbol{q}) \equiv \int e^{i\boldsymbol{q}\boldsymbol{r}} \rho(\boldsymbol{r}) d^3 r. \qquad (11)$$

For the spherical density distribution $\rho(r)$, the form factor is also spherical and can be expressed in the standard way as $F(q) = \int j_0(qr) \rho(r) r^2 dr$.

Good approximation to typical nuclear form factors is given by the Helm model [25–30], where nucleonic density is approximated by a convolution of a sharp-surface density of radius $R_0$ with a Gaussian smoothing profile, i.e,

$$\rho^{(\text{H})}(\boldsymbol{r}) = \int d^3\boldsymbol{r}' \rho_0 \Theta(R_0 - |\boldsymbol{r}'|) \frac{\exp\left(-\frac{(\boldsymbol{r}-\boldsymbol{r}')^2}{2\sigma^2}\right)}{(2\pi)^{3/2}\sigma^3}. \qquad (12)$$

The quantity $R_0$ in Eq. (12) is called the diffraction (box equivalent) radius, and the folding width $\sigma$ represents the surface thickness. The diffraction radius, $R_0$, can be deduced from the first zero, $q_1$, of the microscopic form factor $F(q)$:

$$R_0 = 4.49341/q_1, \qquad (13)$$

while the surface thickness parameter, $\sigma$, can be computed by comparing the values of microscopic and of Helm form factors, $F(q_m)$ and $F^{(\text{H})}(q_m)$, at the first maximum $q_m$ of $F(q)$:

$$\sigma^2 = \frac{2}{q_m^2} \ln \frac{3 R_0^2 j_1(q_m R_0)}{R_0 q_m F(q_m)}. \qquad (14)$$

The geometric radius of the Helm model can be easily computed as

$$R_{\text{Helm}} = \sqrt{\frac{5}{3}} R_{\text{rms}}^{(\text{H})} = \sqrt{(R_0^2 + 5\sigma^2)}. \qquad (15a)$$

From this relation one sees that the geometric radius becomes the box-equivalent radius in the limit of a small surface thickness. The Helm model follows the exact density distribution over a wide range of densities, but some deviations may build up far outside the nucleus at very low densities. Thus the Helm radius is a good approximation to the true geometric radius in well-bound nuclei. The situation changes if one goes towards the drip line where the nucleons become less bound and the outer tail of the density makes a non-negligible contribution to the geometric radius. Since the outer tail is not contained at all in the Helm model, the halo size can be characterized by the difference of these two radii [24], i.e.,

$$\delta R_{\text{halo}} \equiv R_{\text{geom}} - R_{\text{Helm}}. \quad (16)$$

In Ref. [24] it has been shown that $\delta R_{\text{halo}}$ is small in well-bound nuclei, but it becomes enhanced for heavy exotic systems with low neutron separation energies. Furthermore, it was noticed that the halo parameter obtained in the HFB+SLy4 model with DDDI is significantly larger than that predicted in the HFB+SkP model, and in relativistic mean-field models with the finite-range Gogny pairing. In the following, we shall investigate the influence of pairing on $\delta R_{\text{halo}}$.

The bottom portion of Fig. 6 shows the average neutron pairing gaps [19]

$$\langle \Delta_n \rangle = -\frac{1}{N} \int d^3\boldsymbol{r} d^3\boldsymbol{r}' \sum_{\sigma\sigma'} \tilde{h}_n(\boldsymbol{r}\sigma, \boldsymbol{r}'\sigma') \rho_n(\boldsymbol{r}'\sigma', \boldsymbol{r}\sigma). \quad (17)$$

For $\alpha=1/6$, pairing correlations are so strong that they give rise to a non-zero static pairing in the magic nucleus $^{132}$Sn. For larger values of $\alpha$, pairing gaps at $50 \leq N \leq 100$ are almost independent of $\alpha$, while a weak dependence is seen only near the drip line.

The impact of the pairing Hamiltonian on the two-neutron separation energies in the Sn isotopes is illustrated in Fig. 6, middle portion. The most striking result is that the strong pairing at low densities dramatically reduces the shell effect at $N=82$. The presence of this effect gives a strong argument against taking small values of $\alpha$ when aiming at realistic calculations. Another consequence of enhanced pairing is the shift in the position of the two-neutron drip line, $S_{2n}=0$. While for $\alpha=1$ and $1/2$ the last two-neutron bound Sn isotope is calculated at $N=120$; at $\alpha=1/6$ the nuclear binding is increased by pairing and the drip line shifts to $N=126$.

Finally, the top panel of Fig. 6 displays the neutron halo parameter (16). As already pointed out in Ref. [24], this quantity is very small in well-bound nuclei and increases with decreased separation energy. In the case of Sn isotopes considered, there is a gradual rise of $\delta R_{\text{halo}}$ for $N>132$. However, halo size is a very sensitive function of the pairing interaction. Indeed, when decreasing the pairing exponent from $\alpha=1$ to $\alpha=1/6$, the halo parameter increases by an order of magnitude. (Actually, for $\alpha=1/6$, $\delta R_{\text{halo}}$ is nonzero also for well-bound systems.) This is a consequence of a strong feedback between p-h and p-p densities (cf. discussion around Fig. 2).

## V. CONCLUSIONS

This work contains the theoretical analysis of particle and pairing densities in neutron-rich nuclei and their dependence on the choice of pairing interaction. The main goal was to see the effect of pairing on spatial characteristics of nucleonic densities of nuclei far from stability, where the closeness of the particle continuum qualitatively changes the physical situation.

The main conclusion of our work is that, due to the self-consistent feedback between particle and pairing densities, the size of the neutron halo is strongly influenced by pairing correlations; hence by the pairing parametrization assumed. Consequently, experimental studies of neutron distributions in nuclei are extremely important for determining the density dependence of pairing interaction in nuclei. Our analysis suggests that the strong low-density dependence of the pairing force, simulated by taking very small values of $\alpha$ in DDDI, is unphysical. Therefore, predictions of very large halos in drip-line nuclei, obtained with $\alpha=1/6$ [24], should probably be revisited when more reliable information on pairing forces becomes available. The present experimental data are consistent with $1/2 \lesssim \alpha \lesssim 1$. In this context, it is interesting to note that excellent fits to the data were obtained in Refs. [15,16] by taking $\alpha=2/3$. However, at present there is no theoretical argument why the density dependence should be even taken in a form of the power law. Clearly, a more precise determination of the density dependence (including the isovector dependence) is a challenging task for future theoretical work.

## ACKNOWLEDGMENTS

The authors would like to thank Jerry D. Garrett for many inspirational discussions on nuclear structure, nuclear pairing, and exotic nuclei. Jerry devoted a significant part of his scientific life to the question of pairing correlations and their manifestations in nuclear spectroscopy [31,32] and to new exciting physics brought by the development of radioactive beams [33]. Jerry was our close collaborator and a wonderful friend. We miss him dearly.

This research was supported in part by the U.S. Department of Energy under Contract Nos. DE-FG02-96ER40963 (University of Tennessee), DE-FG05-87ER40361 (Joint Institute for Heavy Ion Research), DE-AC05-00OR22725 with UT-Battelle, LLC (Oak Ridge National Laboratory), the Polish Committee for Scientific Research (KBN), and by the Bundesministerium für Bildung und Forschung (BMBF), Project No. 06 ER 808. We thank the Institute for Nuclear Theory at the University of Washington for its hospitality and the Department of Energy for partial support during the completion of this work.

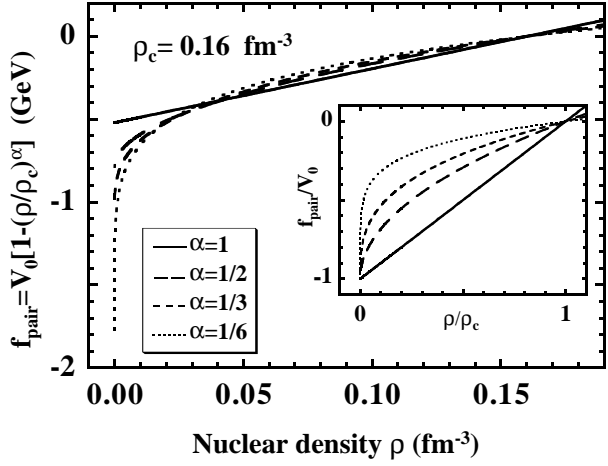

FIG. 1. Radial strength factor $f_{\rm pair}$ of the density-dependent delta interaction, Eq. (9), as a function of $\rho$ for several values of $\alpha$. The value of $\rho_0$ was assumed to be $0.16\,{\rm fm}^{-3}$. At each value of $\alpha$, the strength $V_0$ was adjusted to reproduce the neutron pairing gap in $^{120}$Sn. The inset shows $f_{\rm pair}/|V_0|$ as a function of dimensionless normalized density $\rho_{\rm IS}/\rho_c$.

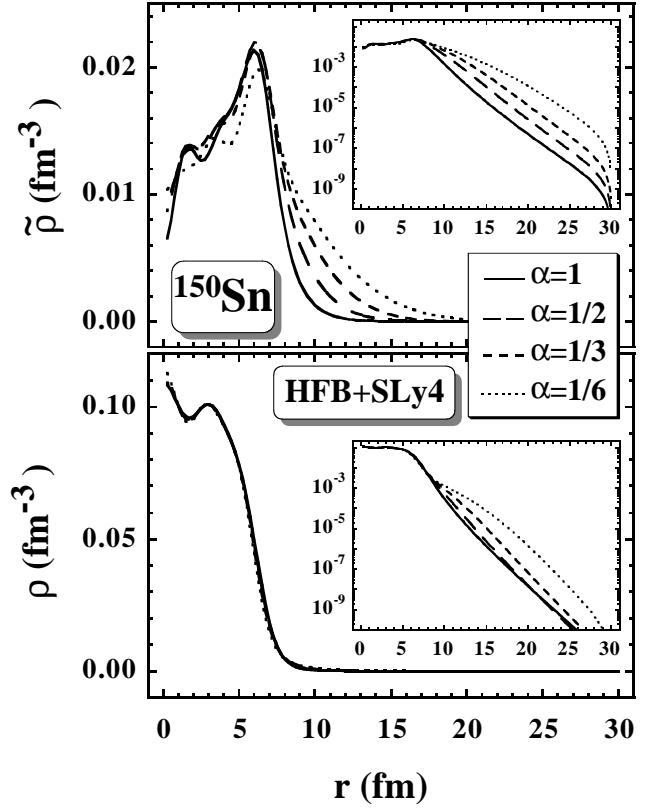

FIG. 2. Self-consistent spherical HFB+SLy4 local densities $\rho(r)$ (top) and $\tilde{\rho}(r)$ (bottom) for neutrons in $^{150}$Sn for several values of $\alpha$. The insets show the same data in logarithmic scale.

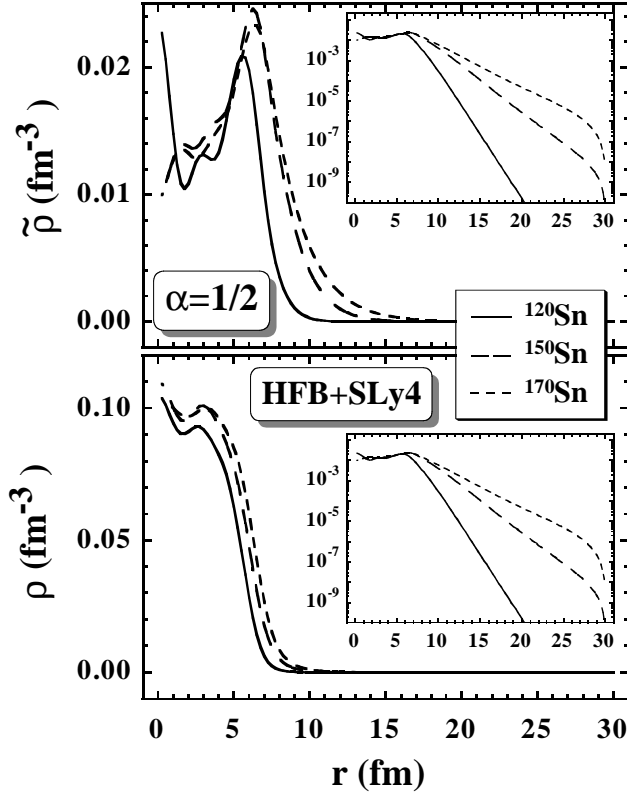

FIG. 3. Self-consistent spherical HFB+SLy4 local densities $\rho(r)$ (top) and $\tilde{\rho}(r)$ (bottom) for neutrons in $^{120,150,170}$Sn and $\alpha=1/2$. The insets show the same data in logarithmic scale. The dramatic fall-off of densities at 30 fm is due to the box boundary conditions.

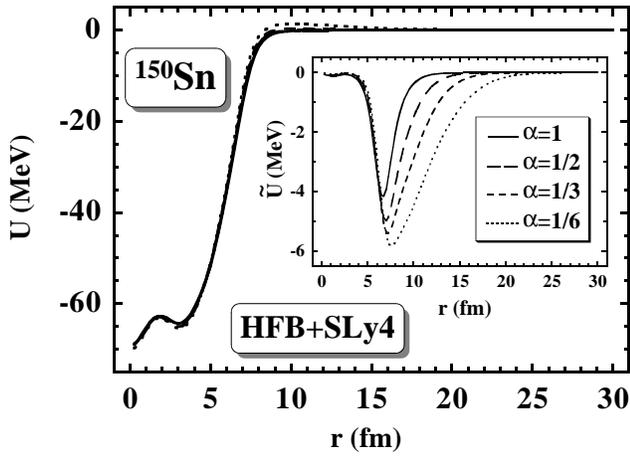

FIG. 4. Self-consistent spherical HFB+SLy4 local potentials $U(r)$ and $\tilde{U}(r)$ (shown in the inset) for neutrons in $^{150}$Sn for several values of $\alpha$.

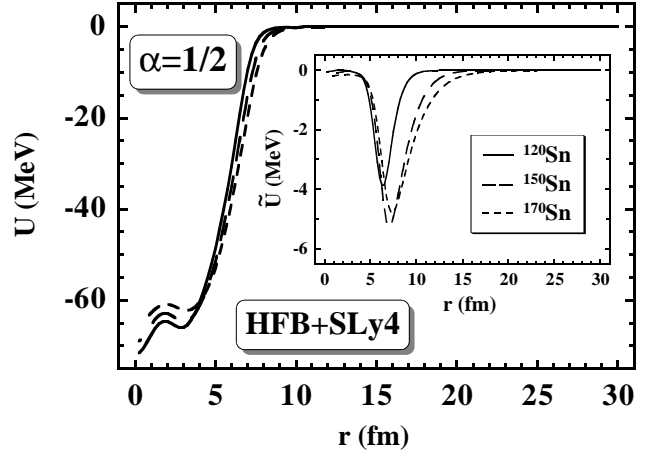

FIG. 5. Self-consistent spherical HFB+SLy4 local potentials $U(r)$ and $\tilde{U}(r)$ (shown in the inset) for neutrons in $^{120,150,170}$Sn and $\alpha=1/2$.

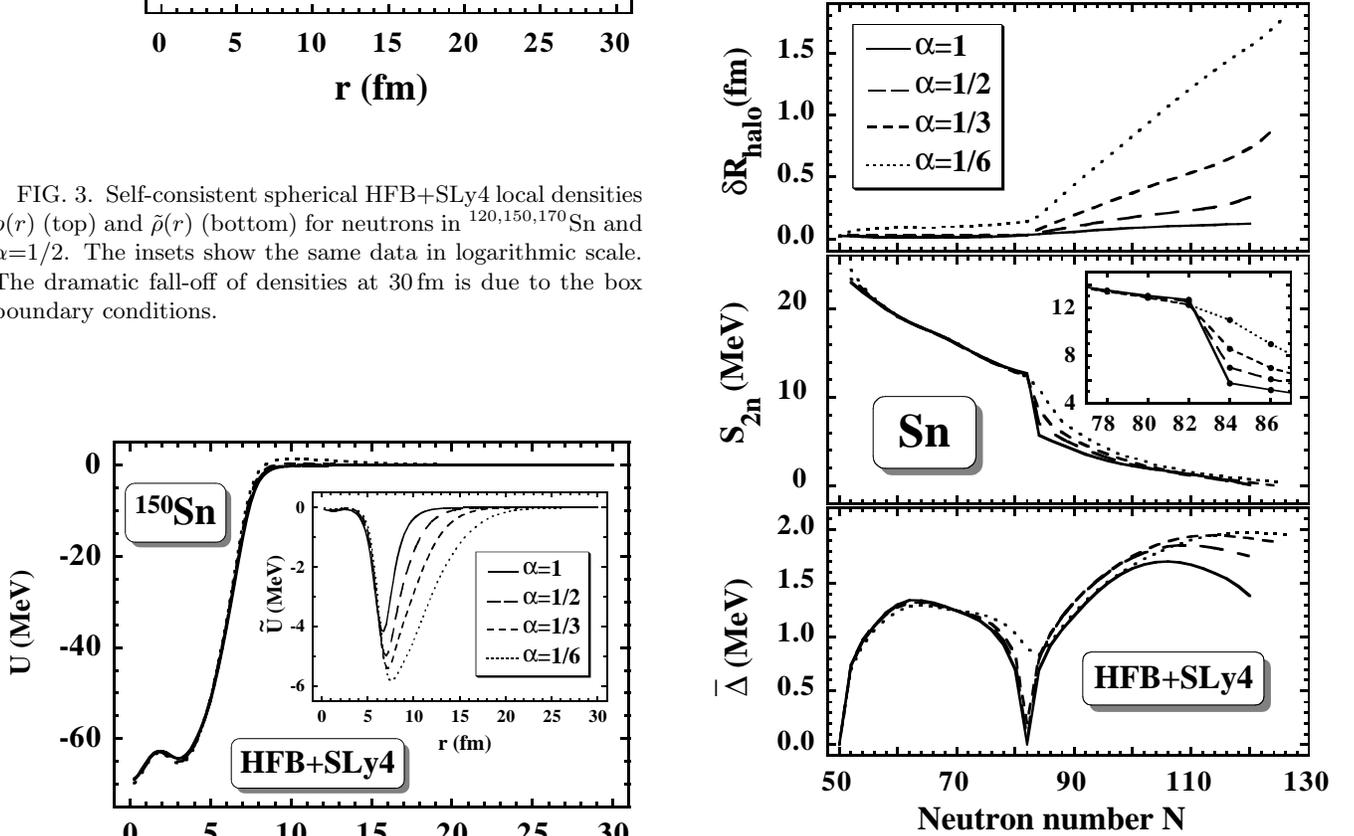

FIG. 6. Neutron halo parameters (top), two-neutron separation energies (middle), and average neutron pairing gaps (17) calculated in the HFB+SLy4 model with different density-dependent pairing interactions (8).